\documentclass[aps,prl,superscriptaddress,showpacs,floatfix,nobibnotes]{revtex4}

\usepackage{graphicx}
\usepackage{longtable}
\usepackage{CJK}
\usepackage{color}

\usepackage{mathptmx, courier, pifont}
\usepackage[scaled=0.92]{helvet}
\usepackage[T1]{fontenc}
\usepackage{textcomp}

\begin{document}

\begin{CJK*}{GB}{}

\title{Exact solution of the two-axis countertwisting Hamiltonian\\
\vskip .5cm
{\small Feng Pan$^{1,3}$, Yao-Zhong Zhang$^{2,4}$~\footnote{Correspondence to: 
School of Mathematics and Physics, The University of Queensland, Brisbane, Qld 4072, Australia. Email address: yzz@maths.uq.edu.au.}, 
and Jerry P. Draayer$^{3}$}}

\address{Department of Physics, Liaoning Normal University, Dalian 116029, P. R. China\\
$^{2}$School of Mathematics and Physics, The University of Queensland, Brisbane, Qld 4072, Australia\\
$^{3}$Department of Physics and Astronomy, Louisiana State University, Baton Rouge, Louisiana 70803-4001, USA\\
$^{4}$CAS Key Laboratory of Theoretical Physics, Institute of Theoretical Physics\\
      Chinese Academy of Sciences, Beijing 100190, P. R. China
}

\date{\today}

\begin{abstract}
It is shown that the two-axis countertwisting Hamiltonian is exactly solvable
when the quantum number of the total angular momentum of the system is an integer
after the Jordan-Schwinger (differential) boson realization of the SU(2) algebra.
Algebraic Bethe ansatz is used to get the exact solution with the help of
the SU(1,1) algebraic structure, from which a set of Bethe ansatz equations
of the problem is derived. It is shown that solutions of the Bethe ansatz equations
can be obtained as zeros of the Heine-Stieltjes polynomials.
The total number of the four sets of the zeros equals exactly to $2J+1$ for a given integer
angular momentum quantum number $J$, which proves the completeness of the solutions.
It is also shown that double degeneracy in level energies may also
occur in the $J\rightarrow\infty$ limit for integer $J$ case except
a unique non-degenerate level with zero excitation energy.

\end{abstract}


\maketitle

\end{CJK*}

\begin{center}
\vskip.2cm\textbf{I. Introduction}
\end{center}\vskip.2cm

Squeezed spin states of both Bose and Fermi many-body
systems~\cite{1,11,12,13,14,15,17,18}, where a component of the
total angular momentum of an ensemble of spins has
less uncertainty~\cite{21,22} than other cases without quantum
mechanical correlations,
have been attracting great attention~\cite{31,32,33,34}, not only
because they are intrinsically interesting, but also
because of being practically useful in precision measurements
\cite{11}, quantum information, and fundamental tests of quantum
mechanics \cite{41}. As shown in \cite{1}, maximal
squeezed spin states of a many-particle system can be generated by using the
two-axis countertwisting mechanism, of which the Hamiltonian
of the system is referred to as the two-axis countertwisting Hamiltonian.
When the number of particles is small, the Hamiltonian can easily be
diagonalized for a given quantum number of the total angular momentum of the system.
However, one needs to handle a huge sparse matrix
when the system contains an ensemble of a large number of particles~\cite{14,15,17,18,34}.
Exact analytical solution to the problem should be helpful, especially when one deals with
a large number of particles.  As noted in \cite{bha}, up till now,
there has been no general analytic solution available,
though there were a few analytic treatments~\cite{51,52,53,54} for a system with
small number of particles.

\vskip .3cm
In fact, significant progresses have been made in finding exact solutions of many-spin systems
since the work of Bethe, Gaudin and Richardson~\cite{be,gau,ric1,ric2}.
Particularly, the Lipkin-Meshkov-Glick (LMG) model, which can be
expressed in terms of the total angular momentum operators of the system
up to their quadratic form,
has been analytically solved by using the algebraic Bethe ansatz~\cite{pan, mori}.
The same problem can also be solved by using the Dyson boson realization of the
SU(2) algebra~\cite{vi1,vi2,zhang},
of which the solutions may be obtained from the Riccati differential equations~\cite{vi1,vi2}.
Discrete phase analysis of the model with applications to
spin squeezing and entanglement was studied in \cite{mar}.
In \cite{pan1}, it was shown that
asymmetric rotor Hamiltonian can also be solved analytically by using
the algebraic Bethe ansatz. However, though the two-axis countertwisting Hamiltonian
is equivalent to a special case of the LMG model~\cite{vi1,vi2} after an Euler rotation, the
procedures used in \cite{pan,mori,pan1} can not be applied
to the two-axis countertwisting Hamiltonian directly.

\vskip .3cm
In this work, we show that the two-axis countertwisting Hamiltonian
is indeed exactly solvable when the quantum number of the total angular momentum
of the system is an integer after the Jordan-Schwinger (differential) boson realization
of the SU(2) algebra.
Similar to \cite{pan1}, exact solution to the problem will be derived based on
the SU(1,1) algebraic structure after suitable transformations.
Moreover, it is shown that solutions of the Bethe ansatz equations
can be obtained from zeros of the Heine-Stieltjes polynomials, which, in turn,
verifies the completeness of the solutions.

\newpage
 \begin{center}
\vskip.2cm\textbf{II. The two-axis countertwisting Hamiltonian}
\end{center}\vskip.2cm

 The two-axis countertwisting Hamiltonian may be written
 as~\cite{1}

 \begin{equation}\label{1}
 {H}_{\rm TA}={\chi\over{2i}}({J}^{2}_{+}-{J}^{2}_{-}),
 \end{equation}
where ${J}_{\pm}$ are the angular momentum raising and lowering
operators, $i=\sqrt{-1}$, and $\chi$ is a constant.
The Hamiltoanian (\ref{1}) is invariant under both parity and
time reversal transformations, namely, it is  $PT$-symmetric.
Due to time-reversal symmetry, similar to the asymmetric rotor case ~\cite{pan1},
level energies of the system
are all doubly degenerate when the quantum number of the
total angular momentum is  a half-integer.
(\ref{1}) is also equivalent to a special LMG Hamiltonian after rotation
of the system by $\pi/4$ around $z$ axis, of which the thermodynamic limit was studied in \cite{vi1,vi2}
by using the Dyson boson (differential) realization and the corresponding
Riccati differential equations.

\vskip .3cm
Using the Jordan-Schwinger realization of SU(2), we have

 \begin{equation}\label{2}
 {J}_{+}=a^{\dagger}b,~~{J}_{-}=b^{\dagger}a, ~~{J}_{0}={1\over{2}}(a^{\dagger}a-b^{\dagger}b),
 \end{equation}
where $a,~b$ and $a^{\dagger},~b^{\dagger}$ are boson annihilation and creation operators
introduced.
It can be observed  that eigenstates of (\ref{1}) may be
expressed as
\begin{equation}
\vert N, \zeta\rangle=F^{(\zeta)}_{t}(a^{\dagger 2},b^{\dagger 2})\vert \nu_{a},\nu_{b}\rangle
\end{equation}
after the Jordan-Schwinger realization,
where $F^{(\zeta)}_{t}(a^{\dagger 2},b^{\dagger 2})$ is a homogenous
polynomial of degree $t$ with variables $\{a^{\dagger 2}, b^{\dagger 2}\}$, $\vert \nu_{a},\nu_{b}\rangle$
is the boson pairing vacuum satisfying $a^{2}\vert \nu_{a},\nu_{b}\rangle=0$,
$b^{2}\vert \nu_{a},\nu_{b}\rangle=0$, in which $\vert \nu_{a},\nu_{b}\rangle
=a^{\dagger\nu_{a}}b^{\dagger\nu_{b}}\vert 0\rangle$ with the boson seniority
numbers $\nu_{a}$ and $\nu_{b}=0$ or $1$, and $N=2J=2t+\nu_{a}+\nu_{b}$,
in which $J$ is the quantum number of the total angular momentum of the system.

\vskip .3cm
Then, we map the boson operators into the Bargmann variables with
$a^{\dagger}\Leftrightarrow x$, $a\Leftrightarrow \partial/\partial x$,
$b^{\dagger}\Leftrightarrow y$, $b\Leftrightarrow \partial/\partial y$.
After this differential realization, however, there are
two parts in the wavefunction. One is the collective part denoted as $F(x,y)$, which is a polynomial
with even powers of $x$ and $y$. Another part denoted as $\Psi_{\rm in}(x,y)$, which is a constant
or proportional to $x$ or $y$.
Thus, the total wavefunction in the Bargmann space may be expressed as $F(x,y)\Psi_{\rm in}(x,y)$.
As a consequence, the Hamiltonian (1) can be expressed as

 \begin{equation}\label{TA}
 {H}_{\rm TA}={\chi\over{2i}}\left(x^2{\partial^2\over{\partial y^2}} -y^2{\partial^2\over{\partial x^2}}
 +2x^2{\partial\over{\partial y}} ({\partial\over{\partial y}})_{\rm in}-
 2y^2{\partial\over{\partial x}} ({\partial\over{\partial x}})_{\rm in}
 \right),
 \end{equation}
where $({\partial\over{\partial y}})_{\rm in}$ and $({\partial\over{\partial y}})_{\rm in}$
indicate that the derivatives are carried out for the intrinsic part $\Psi_{\rm in}(x,y)$ only, while other
derivatives are carried out for the collective part $F(x,y)$ only.
Because the collective part $F(x,y)$ is a function of $x^2$ and $y^2$,
we can make the following transformation:

\begin{equation}\label{z}
 z_{1}=x^2,~~z_{2}=y^2.
 \end{equation}
We also have
\begin{equation}
 {\partial\over{\partial x}}=2x{\partial\over{\partial z_{1}}},~~{\partial\over{\partial y}}=2y{\partial\over{\partial z_{2}}},
 \end{equation}
and
\begin{equation}\label{zz}
 {\partial^2\over{\partial x^2}}=2{\partial\over{\partial z_{1}}}+4z_{1}{\partial^2\over{\partial z_{1}^2}},~~
 {\partial^2\over{\partial y^2}}=2{\partial\over{\partial z_{2}}}+4z_{2}{\partial^2\over{\partial z_{2}^2}}.
 \end{equation}
Substituting (\ref{z}) - (\ref{zz}) into (\ref{TA}) and after separating the collective part from the
intrinsic part,  we get
\begin{eqnarray}\label{TA1}
 {H}_{\rm TA}={\chi\over{i}}\left(
 (1+2\delta_{\hat{\nu}_{b}1})
 z_{1}{\partial\over{\partial z_{2}}}-(1+2\delta_{\hat{\nu}_{a}1})z_{2}{\partial\over{\partial z_{1}}}
 +2z_{1}z_{2}({\partial^2\over{\partial z_{2}^2}}-{\partial^2\over{\partial z_{1}^2}})\right),
 \end{eqnarray}
where $\hat{\nu}_{a}$ and $\hat{\nu}_{b}$ are seniority number operator of $a$- and $b$-bosons,
respectively.
(\ref{TA1}) is thus realized within the new two-dimensional Bargmann space with variables $\{z_{1},z_{2}\}$,
while the intrinsic part characterized by the seniority numbers $\nu_{a}$ and $\nu_{b}$
only affects the first two terms in the Hamiltonian (\ref{TA1}).

\vskip .3cm
By mapping the Bargmann variables $\{z_{1},~z_{2}\}$ to new boson operators with
$z_{1}\Leftrightarrow c^{\dagger}$, $\partial/\partial z_{1}\Leftrightarrow c$,
$z_{2}\Leftrightarrow d^{\dagger}$, $\partial/\partial z_{2}\Leftrightarrow d$,
 Eq. (\ref{TA1}) may  be written as
 \begin{equation}\label{TA2}
 {H}_{\rm TA}={\chi\over{i}}\left( (1+2\delta_{\hat{\nu}_{b}1})  c^{\dagger}d-(1+2\delta_{\hat{\nu}_{a}1})d^{\dagger}c
 +2c^{\dagger}d^{\dagger}(d^{2}-c^{2}) \right).
 \end{equation}

The method outlined below works for integer $J$ case, but may not be applied to half-integer $J$ case directly.
In fact, when $\hat{\nu}_{a}=\hat{\nu}_{b}=\hat{\nu}$, the Hamiltonian (\ref{TA2}) can be expressed
 in terms of two canonical orthonormal boson modes

 \begin{equation}\label{a}
 a^{\dagger}_{1}=\sqrt{1\over{2}}(c^{\dagger}+i d^{\dagger}),~~
 a^{\dagger}_{2}=\sqrt{1\over{2}}(c^{\dagger}-i d^{\dagger}),
 \end{equation}
with
\begin{equation}\label{TA3}
 {H}_{\rm TA}={\chi}\left( (1+2\delta_{\hat{\nu} 1})( a^{\dagger}_{1}a_{1}-a^{\dagger}_{2}a_{2})
 +(a^{\dagger 2}_{1}-a^{\dagger 2}_{2})(a^{2}_{1}+a^{2}_{2})\right).
 \end{equation}
Though (\ref{TA3}) is non-Hermitian, its eigenvalues are all real, mainly
because of its equivalence to the original Hamiltonian (\ref{1}) for this case.
Since $\nu_{a}=\nu_{b}=\nu$, the total angular momentum of the system
should be integer in this case with $J=0,~1,~2,~\cdots$. It will be shown in the following
that (11) can be solved analytically.

\vskip .4cm
\begin{center}
\vskip.2cm\textbf{III. Exact solution for integer $J$ cases}
\end{center}\vskip.2cm

In order to diagonalize (\ref{TA3}), let us introduce two copies of SU(1,1) algebra
generated by $\{S_{+}(1)={1\over{2}}a^{\dagger 2}_{1}$,
$S_{-}(1)={1\over{2}}a^{2}_{1}$,
$S_{0}(1)={1\over{2}}(a^{\dagger}_{1}a_{1}+{1\over{2}})\}$
and $\{S_{+}(2)={1\over{2}}a^{\dagger 2}_{2},~
S_{-}(2)={1\over{2}}a^{2}_{2},~S_{0}(2)={1\over{2}}(a^{\dagger}_{2}a_{2}+{1\over{2}})\}$, which
satisfy the following commutation relations:

\begin{equation}\label{su11}
[S_{0}(l),~ S_{\pm}(j)]=\delta_{lj}S_{\pm}(j),~~~
[S_{+}(l),~ S_{-}(j)]=-\delta_{lj}2S_{0}(j).
\end{equation}
Then, (\ref{TA3}) can be written as
\begin{eqnarray}\label{TA4}
 {H}_{\rm TA}=2\chi\left( (1+2\delta_{\hat{\nu} 1})(S_{0}(1)-S_{0}(2)) + 2(S_{+}(1)-S_{+}(2))(S_{-}(1)+S_{-}(2))
 \right),
 \end{eqnarray}
which can be diagonalized under the Bethe ansatz

\begin{eqnarray}\label{BA}
\vert k,n_{1},n_{2},\nu;\zeta\rangle= S_{+}(w^{(\zeta)}_{1})S_{+}(w^{(\zeta)}_{2})\cdots S_{+}(w^{(\zeta)}_{k})\vert n_{1},n_{2},\nu\rangle
\end{eqnarray}
with $J=2k+n_{1}+n_{2}+\nu$, where
$\vert n_{1},n_{2},\nu\rangle$ is the lowest weight state of the SU$^{(l)}$(1,1) for $l=1,~2$,
satisfying $S_{-}(l)\vert n_{1},n_{2},\nu\rangle=0$ and $2S_{0}(l)\vert n_{1},n_{2},\nu\rangle=
\left(n_{l}+{1\over{2}}\right)\vert n_{1},n_{2},\nu\rangle$ with $n_{l}=0$ or $1$,
and

\begin{equation}\label{S}
S_{+}(w)= {1\over{1-w}}S_{+}(1)+{1\over{1+w}}S_{+}(2).
\end{equation}
By using the commutation relations (\ref{su11}), it can be proven that
\begin{eqnarray}\label{c1}
[S_{0}(1)-S_{0}(2),~S_{+}(w)]= {1\over{1-w}}S_{+}(1)-{1\over{1+w}}S_{+}(2)=S_{+}+w S_{+}(w),
\end{eqnarray}
where  $S_{+}=S_{+}(1)-S_{+}(2)$,

\begin{eqnarray}\label{c2}
[S_{-}(1)+S_{-}(2),~S_{+}(w)]=\Lambda_{0}(w)= {2S_{0}(1)\over{1-w}}+{2S_{0}(2)\over{1+w}}.
\end{eqnarray}

\begin{eqnarray}\label{c3}
S_{+}(w_{1},w_{2})=[[S_{-}(1)+S_{-}(2),~S_{+}(w_{1})],~S_{+}(w_{2})]= {2w_{1}\over{w_{1}-w_{2}}}
S_{+}(w_{1})-{2w_{2}\over{w_{1}-w_{2}}}S_{+}(w_{2}).
\end{eqnarray}

Using Eqs. (\ref{c1}) -(\ref{c3}), we can directly check that

\begin{eqnarray}\label{10}\nonumber
(S_{0}(1)-S_{0}(2))\vert k,n_{1},n_{2},\nu;\zeta\rangle &=&
\left(S_{+}+w^{(\zeta)}_{1}S_{+}(w^{(\zeta)}_{1})\right)
S_{+}(w^{(\zeta)}_{2})\cdots S_{+}(w^{(\zeta)}_{k})\vert n_{1},n_{2},\nu\rangle\\
& &+\cdots+S_{+}(w^{(\zeta)}_{1})\cdots S_{+}(w^{(\zeta)}_{k-1})\left(
S_{+}+w^{(\zeta)}_{k}S_{+}(w^{(\zeta)}_{k})\right)
\vert n_{1},n_{2},\nu\rangle,
\end{eqnarray}
and
\begin{eqnarray}\label{11}\nonumber
S_{+}(S_{-}(1)+S_{-}(2))\vert k,n_{1},n_{2},\nu;\zeta\rangle&=&S_{+}(\left(
\overline{\Lambda}_{0}(w^{(\zeta)}_{1})S_{+}(w_{2}^{(\zeta)})\cdots S_{+}(w^{(\zeta)}_{k})\right.\\
& &+\left.\cdots+S_{+}(w^{(\zeta)}_{1})\cdots S_{+}(w^{(\zeta)}_{k-1})\overline{\Lambda}_{0}(w^{(\zeta)}_{k})\right)\vert n_{1},n_{2},\nu\rangle \nonumber \\
& &+S_{+}\left(S_{+}(w^{(\zeta)}_{1},~w^{(\zeta)}_{2})S_{+}(w^{(\zeta)}_{3})\cdots S_{+}(w^{(\zeta)}_{k})\right.\nonumber \\
& &+S_{+}(w^{(\zeta)}_{1},~w^{(\zeta)}_{3})S_{+}(w^{(\zeta)}_{2})S_{+}(w^{(\zeta)}_{4})\cdots S_{+}(w^{(\zeta)}_{k})\nonumber \\
& &+\cdots+S_{+}(w^{(\zeta)}_{1},~w^{(\zeta)}_{k})S_{+}(w^{(\zeta)}_{2})\cdots S_{+}(w^{(\zeta)}_{k-1})\nonumber \\
& &+\cdots+S_{+}(w^{(\zeta)}_{k},~w^{(\zeta)}_{1})S_{+}(w^{(\zeta)}_{2})\cdots S_{+}(w^{(\zeta)}_{k-1})\nonumber \\
& &+S_{+}(w^{(\zeta)}_{k},~w^{(\zeta)}_{2})S_{+}(w^{(\zeta)}_{3})\cdots S_{+}(w^{(\zeta)}_{k-1})\nonumber \\
& &+\cdots+\left.S_{+}(w^{(\zeta)}_{k},~w^{(\zeta)}_{k-1})S_{+}(w^{(\zeta)}_{2})\cdots S_{+}(w^{(\zeta)}_{k-2})\right)
\vert n_{1},n_{2},\nu\rangle,
\end{eqnarray}
where $\overline{\Lambda}_{0}(w)={n_{1}+{1\over{2}}\over{1-w}}+{n_{2}+{1\over{2}}\over{1+w}}$.

Using Eqs. (\ref{c3}) -(\ref{11}),
one can prove that the eigen-equation ${H}_{\rm TA}\vert k,n_{1},n_{2},\nu;\zeta\rangle=E^{(\zeta)}_{k,n_{1},n_{2},\nu}\vert k,n_{1},n_{2},\nu;\zeta\rangle$
is fulfilled if and only if

\begin{equation}\label{12}
{n_{1}+{1\over{2}}\over{1-w^{(\zeta)}_{l}}}+{n_{2}+{1\over{2}}\over{1+w^{(\zeta)}_{l}}}+{1\over{2}}(1+2\delta_{\nu 1})-
\sum_{j\neq l}{2w^{(\zeta)}_{{j}}\over{w^{(\zeta)}_{l}-w^{(\zeta)}_{j}}}
=0~~{\rm for}~~l=1,2,\cdots,~ k,
\end{equation}
which are independent of the energy scale-factor $\chi$.
The corresponding eigen-energy is given by

\begin{equation}\label{13}
E^{(\zeta)}_{{k,n_{1},n_{2},\nu}}=2\chi(1+2\delta_{\nu 1})
\left(\sum_{l=1}^{k}w_{l}^{(\zeta)}+(n_{1}-n_{2})/2\right)\end{equation}
with $J=2k+n_{1}+n_{2}+\nu$, where $k$ is the number of boson-quartets,
$n_{1}$ and $n_{2}$ are the numbers of two different boson pairs, while $2\nu$ is the total number of
unpaired bosons, in which the bosons are the $a$- and $b$-bosons introduced in (\ref{2}).
It can be inferred from (\ref{13})
that the spectrum of the model  after the Jordan-Schwinger two-boson realization
is generated from the non-linear boson-quartet excitations
based on the single-boson and the boson-pairing excitations, where the single-boson excitation
affects both the scaling of the energy and the boson-quartet excitations, while
the boson-pairing excitation energies contribute to the total energy linearly.
Moreover, as shown previously~\cite{3,4,5}, though the eigenstates provided in (\ref{BA})
are not normalized, they are always orthogonal with
\begin{equation}
 \langle k^{\prime},n^{\prime}_{1},2n^{\prime}_{2},\nu^{\prime};
 \zeta^{\prime}\vert k,n_{1},n_{2},\nu;\zeta\rangle=({\cal N}(k,\zeta;n_{1},n_{2},\nu))^{-2} \delta_{k k^{\prime}}
 \delta_{n_{1} n_{1}^{\prime}} \delta_{n_{2} n_{2}^{\prime}} \delta_{\nu \nu^{\prime}}\delta_{\zeta\zeta^{\prime}},
\end{equation}
where ${\cal N}(k,\zeta;n_{1},n_{2},\nu)$ is the corresponding normalization constant.

\vskip .3cm

In order to find solutions of Eq. (\ref{12}),
one may change variables with $u_{l}=1/w_{l}$.
Then, Eq. (\ref{12}) can be written as

\begin{equation}\label{14}
{n_{1}+{1\over{2}}\over{u_{l}-1}}+{n_{2}+{1\over{2}}\over{u_{l}+1}}+{{1\over{2}}(1+2\delta_{\nu 1})\over{u_{l}}}+
\sum_{j\neq l}{2\over{u_{l}-u_{j}}}
=0~~{\rm for}~~l=1,2,\cdots,~ k.
\end{equation}

According to the Heine-Stieltjes correspondence \cite{3,4,5},  zeros $\{u_{l}\}$
of the Heine-Stieltjes polynomials $y_{k}(u)$
of degree $k$ are roots of
 Eq. (\ref{14}), where $y_{k}(u)$ should satisfy
the following second-order Fuchsian equation:
\begin{equation}\label{15}
A(u)y_{k}^{\prime\prime}(u)+B(u)y^{\prime}_{k}(u)-V(u)y_{k}(u)=0.
\end{equation}
Here, $A(u)=u(u^2-1)$,
the polynomial  $B(u)$ is given as
\begin{equation}\label{151}
B(u)/A(u)={n_{1}+{1\over{2}}\over{u-1}}+{n_{2}+{1\over{2}}\over{u+1}}+{{1\over{2}}(1+2\delta_{\nu 1})\over{u}},
\end{equation}

\begin{equation}\label{152}
{y_{k}^{\prime\prime}(u_{l})\over{y_{k}^{\prime}(u_{l})}}
=\sum_{1\leq j(\neq l)\leq k}{2\over{u_{l}-u_{j}}},
\end{equation}
and $V(u)$ is a Van Vleck polynomial of degree $1$, which is determined according to Eq. (\ref{15}).
Actually, the polynomial $y_{k}(u)$, of which the zeros satisfy (\ref{14}),
is an extended type of Niven or Lam\'{e} function, which is a special type of Heine-Stieltjes polynomials.
Since $n_{1}+{1\over{2}}$, $n_{2}+{1\over{2}}$, and ${1\over{2}}(1+2\delta_{\nu 1})$ are always real and positive,
zeros of the Heine-Stieltjes polynomial are all real and satisfy the interlacing condition.
Let these zeros be arranged as $u_1< u_2 <\cdots< u_k$, which are in the union of
two open intervals: $\{u_{1},u_{2},\cdots,u_k\}\in (-1,0)\bigcup (0,1)$.
An electrostatic interpretation of the location of
zeros of $y_{k}(u)$ may be stated as follows.
Put three positive fixed fractional charges ${1\over{2}}n_{2}+{1\over{4}}$, ${1\over{4}}(1+2\delta_{\nu 1})$,
and ${1\over{2}}n_{1}+{1\over{4}}$ at $-1$, $0$, and $+1$ along a real line, respectively,
and allow $k$ positive unit charges to move freely along the real line
under such situation.
There are $k+1$ different configurations for the position
of these $k$ charges $\{u^{(\zeta)}_{1},\cdots, u^{(\zeta)}_k\}$
with $\zeta=1,2,\cdots, k+1$,
corresponding to global minimums of the total electrostatic energy of the system~\cite{3}.
It follows from this that the total number of these configurations
is exactly the number of ways to put the $k$ zeros into the two open intervals, which is $k+1$.
Thus, there are $k+1$ different polynomials $y_{k}(u)$ for given $\{n_{1},n_{2},\nu\}$.
Since $0\leq n_{1},n_{2},\nu\leq 1$,
for a given integer $J$, there are four different cases.
Specifically, when $J$ is a fixed even integer, there are $k+1$ solutions with $J=2k$ and $\{n_{1}=n_{2}=\nu=0\}$,
while there are $k$ solutions for cases with $\{n_{1}=n_{2}=1,\nu=0\}$,
or $\{n_{1}=\nu=1,n_{2}=0\}$, or $\{n_{2}=\nu=1,n_{1}=0\}$;
when $J$ is a fixed odd integer, there are $k+1$ solutions when $J=2k+1$ with $\{n_{1}=1,n_{2}=\nu=0\}$, or $\{n_{2}=1,n_{1}=\nu=0\}$,
or $\{n_{1}=n_{2}=0,\nu=1\}$,
while there are $k$ solutions for the case with $\{n_{1}=n_{2}=\nu=1\}$.
It is obvious that the total number of different solutions equals exactly to $2J+1$
for both even and odd $J$ cases, which proves the completeness of
the solutions provided by (\ref{14}) for the Hamiltonian (\ref{TA4}).
Therefore, for a given $J$, $2J+1$ solutions in this Bethe ansatz approach split into $4$ sets of solutions 
provided by  (\ref{14}) with different $\{n_{1},n_{2},\nu\}$.

\vskip .3cm

Once the Bethe ansatz equations shown in (\ref{12}) are solved,
the eigenstate (\ref{BA}), up to a normalization constant,
can be expressed in terms of the original
$a$- and $b$-boson operators as

\begin{equation}\label{estate}
\vert  k,n_{1},n_{2},\nu;\zeta\rangle=
\sum_{q=0}^{k}\sum_{\rho=0}^{k-q}
\left(\begin{array}{c}
k-q\\
\rho
\end{array}\right)
(-)^{\rho}(2i)^{q}S^{(k,\zeta)}_{q}
a^{\dagger 4k-2q-4\rho} b^{\dagger 2q+4\rho}\vert n_{1},n_{2},\nu\rangle,
\end{equation}
where
\begin{equation}\label{Sq}
S^{(k,\zeta)}_{0}=1,~~S^{(k,\zeta)}_{q\geq1}=\sum_{1\leq\mu_{1}\neq\cdots\neq\mu_{q}\leq k}w^{(\zeta)}_{\mu_{1}}
\cdots w^{(\zeta)}_{\mu_{q}}
\end{equation}
are the symmetric functions of $\{w^{(\zeta)}_{1},\cdots,w^{(\zeta)}_{k}\}$,
which are related to the expansion coefficients of $y_{k}(u)$ when
it is expanded in terms of powers of $u$~\cite{3,4,5}.
Thus, when $J=2k$, we have

\begin{equation}\label{exp1}\small
\vert J=2k,\zeta\rangle=\left\{
\begin{tabular}{c}
$\small\sum_{q=0}^{k}\sum_{\rho=0}^{k-q}
{(k-q)!(-)^{\rho}(2i)^{q}\over{{(k-q-\rho)!\rho!((4k-2q-4\rho)!(2q+4\rho)!)^{-{1\over{2}}} }}}
S_{q}^{(k,\zeta)}\vert J=2k, ~M=2k-2q-4\rho\rangle~~~~~~~~~~~~~~~~~~~~~~~~~~~{\rm for}~~n_{1}=n_{2}=\nu=0$,~~\\\\
$\small\sum_{q=0}^{k-1}\sum_{\rho=0}^{k-q-1}{(k-q)!(-)^{\rho}\over{(k-q-\rho)!\rho!(2i)^{-q}}}
S^{(k-1,\zeta)}_{q}
({\vert J=2k, ~M=2k-2q-4\rho\rangle\over{((4k-2q-4\rho)!(2q+4\rho)!)^{-{1\over{2}}}}}
+
{\vert J=2k, ~M=2k-2q-4\rho-4\rangle\over{((4k-2q-4\rho-4)!(2q+4\rho+4)!)^{-{1\over{2}}}}})~~~~{\rm for}~n_{1}=n_{2}=1,\nu=0$,\\\\
$\small\sum_{q=0}^{k-1}\sum_{\rho=0}^{k-q-1}{(k-q)!(-)^{\rho}\over{(k-q-\rho)!\rho!(2i)^{-q}}}
S^{(k-1,\zeta)}_{q}
({\vert J=2k, ~M=2k-2q-4\rho-1\rangle\over{((4k-2q-4\rho-1)!(2q+4\rho+1)!)^{-{1\over{2}}}}}
+
{i~\vert J=2k, ~M=2k-2q-4\rho-3\rangle\over{((4k-2q-4\rho-3)!(2q+4\rho+3)!)^{-{1\over{2}}}}})
~{\rm for}~n_{1}=\nu=1,n_{2}=0$,\\\\
$\small\sum_{q=0}^{k-1}\sum_{\rho=0}^{k-q-1}{(k-q)!(-)^{\rho}\over{(k-q-\rho)!\rho!(2i)^{-q}}}
S^{(k-1,\zeta)}_{q}
({\vert J=2k, ~M=2k-2q-4\rho-1\rangle\over{((4k-2q-4\rho-1)!(2q+4\rho+1)!)^{-{1\over{2}}}}}
-
{i~\vert J=2k, ~M=2k-2q-4\rho-3\rangle\over{((4k-2q-4\rho-3)!(2q+4\rho+3)!)^{-{1\over{2}}}}})
~{\rm for}~n_{2}=\nu=1,n_{1}=0$.\\\\
\end{tabular}
\right.
\end{equation}
When $J=2k+1$, we have

\begin{equation}\label{exp2}\small
\vert J=2k+1,\zeta\rangle=\left\{
\begin{tabular}{c}
$\small\sum_{q=0}^{k}\sum_{\rho=0}^{k-q}{(k-q)!(-)^{\rho}(2i)^{q}\over{(k-q-\rho)!\rho!
{((4k-2q-4\rho+1)!(2q+4\rho+1)!)^{-{1\over{2}}} }}}
S_{q}^{(k,\zeta)}\vert J=2k+1, ~M=2k-2q-4\rho\rangle~~~~~{\rm for}~~n_{1}=n_{2}=0,\nu=1$,\\\\
$\small\sum_{q=0}^{k}\sum_{\rho=0}^{k-q}{(k-q)!(-)^{\rho}\over{(k-q-\rho)!\rho!(2i)^{-q}}}
S^{(k,\zeta)}_{q}
({\vert J=2k+1, ~M=2k-2q-4\rho+1\rangle\over{((4k-2q-4\rho+2)!(2q+4\rho)!)^{-{1\over{2}}}}}
+
{i~\vert J=2k+1, ~M=2k-2q-4\rho-1\rangle\over{((4k-2q-4\rho)!(2q+4\rho+2)!)^{-{1\over{2}}}}})
~~~{\rm for}~n_{1}=1,n_{2}=
\nu=0$,\\\\
$\small\sum_{q=0}^{k}\sum_{\rho=0}^{k-q}{(k-q)!(-)^{\rho}\over{(k-q-\rho)!\rho!(2i)^{-q}}}
S^{(k,\zeta)}_{q}
({\vert J=2k+1, ~M=2k-2q-4\rho+1\rangle\over{((4k-2q-4\rho+2)!(2q+4\rho)!)^{-{1\over{2}}}}}
-
{i~\vert J=2k+1, ~M=2k-2q-4\rho-1\rangle\over{((4k-2q-4\rho)!(2q+4\rho+2)!)^{-{1\over{2}}}}})
~~~{\rm for}~n_{2}=1,n_{1}=\nu=0$,\\\\
$\small\sum_{q=0}^{k-1}\sum_{\rho=0}^{k-q-1}{(k-q)!(-)^{\rho}\over{(k-q-\rho)!\rho!(2i)^{-q}}}
S^{(k-1,\zeta)}_{q}
({\vert J=2k+1, ~M=2k-2q-4\rho\rangle\over{((4k-2q-4\rho+1)!(2q+4\rho+1)!)^{-{1\over{2}}}}}
+
{\vert J=2k+1, ~M=2k-2q-4\rho-4\rangle\over{((4k-2q-4\rho-3)!(2q+4\rho+5)!)^{-{1\over{2}}}}})
~{\rm for}~n_{1}=n_{2}=\nu=1$.\\\\
\end{tabular}
\right.
\end{equation}

\vskip .3cm
In order to solve (\ref{14}) more easily, as shown
in \cite{3, 4,5,6}
for the extended Heine-Stieltjes polynomials,
one may simply write
\begin{equation}\label{16}
y^{(\zeta)}_{k}(u)=\sum_{j=0}^{k}b^{(\zeta)}_{j}u^{j},
\end{equation}
where $\{b^{(\zeta)}_{j}\}$ ($j=0,~1,~\cdots,~k$) are the $\zeta$-th set of the expansion
coefficients to be determined.
Substitution of (\ref{16}) into (\ref{15}) yields
the condition to determine
the corresponding Van Vleck polynomial with

\begin{equation}\label{17}
V^{(\zeta)}(u)=k({1\over{2}}(1+2\delta_{\nu 1})+n_{1}+n_{2}+k) u+g^{(\zeta)}_{0}.
\end{equation}
The expansion coefficients $b^{(\zeta)}_{j}$ and $g^{(\zeta)}_{0}$ satisfy
the following three-term relations:

\begin{equation}\label{18}
{j}(n_{1}-n_{2})b^{(\zeta)}_{j}-
(k-j+1)(k+j+n_{1}+n_{2}+{1\over{2}}(1+2\delta_{\nu 1})-1)b^{(\zeta)}_{j-1}-
(j+1)({1\over{2}}(1+2\delta_{\nu 1})+j)b_{j+1}=g^{(\zeta)}_{0}b^{(\zeta)}_{j}\end{equation}
with $b^{(\zeta)}_{j}=0$ for $j\leq-1$ or $j\geq k+1$, which is
equivalent to the eigenvalue problem
with

\begin{equation}\label{19}
{\bf F}{\bf b}^{(\zeta)}=g^{(\zeta)}_{0}{\bf b}^{(\zeta)},\end{equation}
where the transpose of  ${\bf b}^{(\zeta)}$ is
related to the expansion coefficients $\{b^{(\zeta)}_{j}\}$ with
$({\bf b^{(\zeta)}})^{\rm T}=\left( b^{(\zeta)}_{0},b^{(\zeta)}_{1},\cdots,b^{(\zeta)}_{k-1},b^{(\zeta)}_{k}\right)$, and
${\bf F}$ is the $(k+1)\times(k+1)$ tridiagonal matrix with entries determined
by (\ref{18}).

\vskip .3cm

In addition, (\ref{16}) can also be written in terms of the zeros
$\{u^{(\zeta)}_{j}\}$ ($j=1,\cdots,k$) of $y^{(\zeta)}_{k}(u)$ with

\begin{equation}\label{20}
y^{(\zeta)}_{k}(u)=\prod_{j=1}^{k}(u-u^{(\zeta)}_{j})=\sum_{q=0}^{k}(-1)^{q} \tilde{S}_{q}^{(k,\zeta)}u^{k-q},
\end{equation}
where  $\tilde{S}_{q}^{(k,\zeta)}$ is the same symmetric function of $\{u^{(\zeta)}_{1},\cdots,u^{(\zeta)}_{k}\}$ as
that of $\{w^{(\zeta)}_{1},\cdots,~w^{(\zeta)}_{k}\}$ given in (\ref{Sq}).
In comparison of (\ref{20}) with (\ref{16}), we get
\begin{equation}\label{21}
b^{(\zeta)}_{k-q}=(-1)^{q}\tilde{S}^{(k,\zeta)}_{q}\end{equation}
when the overall factor of $\{b^{(\zeta)}_{j}\}$ is chosen with
$b^{(\zeta)}_{k}=1~\forall~\zeta$.
Hence, the symmetric function $\tilde{S}_{q}^{(k,\zeta)}$ is known
after the expansion coefficients ${\bf b}^{(\zeta)}$ are obtained according to (\ref{19}).
Then, the symmetric functions $S^{(k,\zeta)}_{q}$ defined in (\ref{Sq})
can be obtained from $\tilde{S}_{q}^{(k,\zeta)}$ with

\begin{equation}\label{22}
S^{(k,\zeta)}_{q}=\tilde{S}_{k-q}^{(k,\zeta)}\prod^{k}_{j=1}w^{(\zeta)}_{j}=(-1)^{q}~b^{(\zeta)}_{q}/b^{(\zeta)}_{0}
\end{equation}
because $u^{(\zeta)}_{j}=1/w^{(\zeta)}_{j}$, which can then be used in the
eigenstates (\ref{exp1}) and (\ref{exp2}) to avoid unnecessary computation of $S^{(k,\zeta)}_{q}$
from $\{w^{(\zeta)}_{1},\cdots,~w^{(\zeta)}_{k}\}$.

\vskip .4cm

\begin{center}
\vskip.2cm\textbf{IV. ~Some numerical examples of the solution}
\end{center}\vskip.2cm

In order to demonstrate the method and solutions outlined previously,
in this section, we provide some examples of the solution of (1) for integer $J$ cases.
Similar to what was shown in \cite{6}, a Wolfram Mathematica package according to
(\ref{16})-(\ref{19}) is compiled, which is very efficient even when $J$ is a large number
due to the fact that to generate and diagonalize a tridiagonal matrix are easier and more CPU time saving
than other more complicated sparse matrices.
When $J\leq 1$, the solutions are trivial with $k=0$, of which
the eigen-energies are simply given by

\begin{equation}\label{23}
E^{(\zeta)}_{ k,n_{1},n_{2},\nu}=\chi(1+2\delta_{\nu 1})
\left(
n_{1}-n_{2}\right)\end{equation}
with $J=2k+n_{1}+n_{2}+\nu$,
while the corresponding eigenstates are given by (\ref{exp1}) and (\ref{exp2})
with $k=0$. When $J=2$, there is only one non-trivial case with
$\{k=1,n_{1}=n_{2}=\nu=0\}$. When $J=3$,
the only trivial case is that with $\{k=0,n_{1}=n_{2}=\nu=1\}$.
When $J\geq4$, all solutions are non-trivial.

\begin{table*}[h]
\caption{The Heine-Stieltjes Polynomials $y^{(\zeta)}_{k}(u)$,
 $g^{(\zeta)}_{0}$ of the corresponding Van Vleck Polynomial $V^{(\zeta)}(u)$,
 and the corresponding eigenenergy $E^{(\zeta)}_{ k,n_{1},n_{2},\nu}/\chi$
 of the Hamiltonian (1) for $J\leq5$, where the order of $\zeta$ is arranged according to
 the value of the eigen-energy of (1) for a given set of $\{ k,n_{1},n_{2},\nu\}$.}
\begin{tabular}{ccccc}
\hline\hline
$J$&$\{k,\zeta; n_{1},n_{2},\nu\}$ &{$y^{(\zeta)}_{k}(u)$} &$g^{(\zeta)}_{0}$ &$E^{(\zeta)}_{ k,n_{1},n_{2},\nu}/\chi$ \\
\hline
\hline
0 &$\{0, 1; 0,0,0\}$ &$1$ &$0$ &$0$\\
1 &$\{0, 1; 1,0,0\}$ &$1$ &$0$ &$1$\\
  &$\{0, 1; 0,0,1\}$ &$1$ &$0$ &$0$\\
  &$\{0, 1; 0,1,0\}$ &$1$ &$0$ &$-1$\\
2 &$\{0, 1; 1,0,1\}$ &$1$&$0$&$3$\\
  &$\{0, 1; 1,1,0\}$ &$1$&$0$&$0$\\
  &$\{0, 1; 0,1,1\}$ &$1$&$0$&$-3$\\
  &$\{1, 1; 0,0,0\}$ &$0.57735+u$&$-0.866025$&$-3.4641$\\
  &$\{1, 2; 0,0,0\}$ &$-0.57735+u$&$0.866025$&$3.4641$\\
3 &$\{0, 1; 1,1,1\}$ &$1$&$0$&$0$\\
  &$\{1, 1; 0,0,1\}$ &$0.774597+u$&$-1.93649$&$-7.74597$\\
  &$\{1, 2; 0,0,1\}$ &$-0.774597+u$&$1.93649$&$7.74597$\\
  &$\{1, 1; 0,1,0\}$ &$0.289898+u$&$-1.72474$&$-7.89898$\\
  &$\{1, 2; 0,1,0\}$ &$-0.689898+u$&$0.724745$&$1.89898$\\
  &$\{1, 1; 1,0,0\}$ &$0.689898+u$&$-0.724745$&$-1.89898$\\
  &$\{1, 2; 1,0,0\}$ &$-0.289898+u$&$1.72474$&$7.89898$\\
4 &$\{2, 1; 0,0,0\}$ &$0.142857 +1.03016 u+u^2$&$1.03016 $&$-14.4222$\\
  &$\{2, 2; 0,0,0\}$ &$-0.6+u^2$&$0$&$0$\\
  &$\{2, 3; 0,0,0\}$ &$0.142857 -1.03016 u+u^2$&$-1.03016$&$14.4222$\\
  &$\{1, 1; 0,1,1\}$ &$0.527202 +u$&$-2.84521$&$-14.3808$\\
  &$\{1, 2; 0,1,1\}$ &$-0.812917+u$&$1.84521 $&$4.38083$\\
  &$\{1, 1; 1,0,1\}$ &$0.812917+u$&$-1.84521 $&$-4.38083$\\
  &$\{1, 2; 1,0,1\}$ &$-0.527202 +u$&$2.84521$&$14.3808$\\
  &$\{1, 1; 1,1,0\}$ &$0.377964 +u$&$-1.32288 $&$-5.2915$\\
  &$\{1, 2; 1,1,0\}$ &$-0.377964  +u$&$1.32288$&$5.2915$\\
5 &$\{2, 1; 0,0,1\}$ &$0.333333 +1.27657 u+u^2$&$1.27657 $&$-22.9783$\\
  &$\{2, 2; 0,0,1\}$ &$-0.714286+u^2$&$0$&$0$\\
  &$\{2, 3; 0,0,1\}$ &$0.333333 -1.27657 u+u^2$&$-1.27657 $&$22.9783$\\

  &$\{2, 1; 0,1,0\}$ &$0.0706856 +0.777127 u+u^2$&$0.777127 $&$-22.9883$\\
  &$\{2, 2; 0,1,0\}$ &$-0.40046-0.347913 u+u^2$&$-0.347913$&$-2.73757$\\
  &$\{2, 3; 0,1,0\}$ &$0.186917 -1.09588 u+u^2,$&$-1.09588$&$10.7259$\\

  &$\{2, 1; 1,0,0\}$ &$0.186917 +1.09588 u+u^2$&$1.09588  $&$-10.7259$\\
  &$\{2, 2; 1,0,0\}$ &$-0.40046+0.347913 u+u^2$&$0.347913$&$2.73757 $\\
  &$\{2, 3; 1,0,0\}$ &$0.0706856 -0.777127 u+u^2$&$-0.777127 $&$22.9883$\\

   &$\{1, 1; 1,1,1\}$ &$0.57735 +u$&$-2.59808$&$-10.3923 $\\
   &$\{1, 2; 1,1,1\}$ &$-0.57735 +u$&$2.59808 $&$10.3923$\\
 \hline\hline
\end{tabular}\label{t1}
\end{table*}

\begin{table*}[h]
\caption{The same as Table \ref{t1}, but for $J=12$.}
\begin{tabular}{ccccc}
\hline\hline
&$\{k,\zeta; n_{1},n_{2},\nu\}$ &{$y^{(\zeta)}_{k}(u)$} &$g^{(\zeta)}_{0}$ &$E^{(\zeta)}_{ k,n_{1},n_{2},\nu}/\chi$ \\
\hline
\hline
&$\{6, 1; 0,0,0\}$ &$0.000541694 +0.0376684 u+0.429523 u^2+1.80288 u^3+3.43433 u^4+3.0234 u^5+u^6$&$0.0376684 $&$-139.076$\\
  &$\{6, 2; 0,0,0\}$ &$-0.00319288-0.123789 u-0.758387 u^2-1.34125 u^3-0.0153776 u^4+1.68567 u^5+u^6$&$-0.123789$&$-77.5408$\\
  &$\{6, 3; 0,0,0\}$ &$0.0165927 +0.24233 u+0.37415 u^2-0.847361 u^3-1.35063 u^4+0.634983 u^5+u^6$&$0.24233 $&$-29.2092$\\
  &$\{6, 4; 0,0,0\}$ &$-0.0497738+0.647059 u^2-1.57143 u^4+u^6$&$0$&$0$\\
  &$\{6, 5; 0,0,0\}$ &$0.0165927 -0.24233 u+0.37415 u^2+0.847361 u^3-1.35063 u^4-0.634983 u^5+u^6$&$-0.24233$&$29.2092$\\
  &$\{6, 6; 0,0,0\}$ &$-0.00319288+0.123789 u-0.758387 u^2+1.34125 u^3-0.0153776 u^4-1.68567 u^5+u^6$&$0.123789 $&$77.5408$\\
  &$\{6, 7; 0,0,0\}$ &$0.000541694 -0.0376684 u+0.429523 u^2-1.80288 u^3+3.43433 u^4-3.0234 u^5+u^6$&$-0.0376684$&$139.076$\\

  &$\{5, 1; 0,1,1\}$ &$0.00637508 +0.144583 u+0.906987 u^2+2.29763 u^3+2.5234 u^4+u^5$&$0.144583$&$-139.076$\\
  &$\{5, 2; 0,1,1\}$ &$-0.027919-0.346847 u-1.01394 u^2-0.483239 u^3+1.18565 u^4+u^5$&$-0.346847 $&$-77.5399$\\
  &$\{5, 3; 0,1,1\}$ &$0.0854689 +0.368501 u-0.238076 u^2-1.29453 u^3+0.127591 u^4+u^5$&$0.368501 $&$-28.8692$\\
  &$\{5, 4; 0,1,1\}$ &$-0.114417+0.232435 u+0.669984 u^2-1.0747 u^3-0.699757 u^4+u^5$&$0.232435 $&$9.18883$\\
  &$\{5, 5; 0,1,1\}$ &$0.0520016 -0.472687 u+0.993523 u^2+0.0512233 u^3-1.62042 u^4+u^5$&$-0.472687$&$51.5391$\\
  &$\{5, 6; 0,1,1\}$ &$-0.0137782+0.252043 u-1.33025 u^2+2.91356 u^3-2.82082 u^4+u^5$&$0.252043 $&$106.758$\\

  &$\{5, 1; 1,0,1\}$ &$0.0137782 +0.252043 u+1.33025 u^2+2.91356 u^3+2.82082 u^4+u^5$&$0.252043 $&$-106.758$\\
  &$\{5, 2; 1,0,1\}$ &$-0.0520016-0.472687 u-0.993523 u^2+0.0512233 u^3+1.62042 u^4+u^5$&$-0.472687 $&$-51.5391$\\
  &$\{5, 3; 1,0,1\}$ &$0.114417 +0.232435 u-0.669984 u^2-1.0747 u^3+0.699757 u^4+u^5$&$0.232435 $&$-9.18883$\\
  &$\{5, 4; 1,0,1\}$ &$-0.0854689+0.368501 u+0.238076 u^2-1.29453 u^3-0.127591 u^4+u^5$&$0.368501$&$28.8692$\\
  &$\{5, 5; 1,0,1\}$ &$0.027919 -0.346847 u+1.01394 u^2-0.483239 u^3-1.18565 u^4+u^5$&$-0.346847$&$77.5399$\\
  &$\{5, 6; 1,0,1\}$ &$-0.00637508+0.144583 u-0.906987 u^2+2.29763 u^3-2.5234 u^4+u^5$&$0.144583 $&$139.076$\\

  &$\{5, 1; 1,1,0\}$ &$0.00133825 +0.0714339 u+0.618781 u^2+1.87815 u^3+2.32082 u^4+u^5$&$0.0714339$&$-106.758$\\
  &$\{5, 2; 1,1,0\}$ &$-0.00736718-0.189934 u-0.724024 u^2-0.383373 u^3+1.12091 u^4+u^5$&$-0.189934 $&$-51.5621$\\
  &$\{5, 3; 1,1,0\}$ &$0.0351315 +0.207975 u-0.233945 u^2-1.03515 u^3+0.257387 u^4+u^5$&$0.207975 $&$-11.8398$\\
  &$\{5, 4; 1,1,0\}$ &$-0.0351315+0.207975 u+0.233945 u^2-1.03515 u^3-0.257387 u^4+u^5$&$0.207975 $&$11.8398$\\
  &$\{5, 5; 1,1,0\}$ &$0.00736718 -0.189934 u+0.724024 u^2-0.383373 u^3-1.12091 u^4+u^5$&$-0.189934$&$51.5621$\\
  &$\{5, 6; 1,1,0\}$ &$-0.00133825+0.0714339 u-0.618781 u^2+1.87815 u^3-2.32082 u^4+u^5$&$0.0714339 $&$106.758$\\

\hline\hline
\end{tabular}\label{t2}
\end{table*}

The Heine-Stieltjes polynomials $y_{k}^{(\zeta)}(u)$
and the corresponding coefficient $g_{0}^{(\zeta)}$ in the Van Vleck polynomial shown in (\ref{17})
up to $J=5$ are shown in Table \ref{t1}, while the $J=12$ case is provided in Table \ref{t2}.
For any case, it can be verified that any zero of $y_{k}^{(\zeta)}(u)$ indeed lies in one of the
intervals $(-1, 0)$ and $(0, 1)$. In addition,
the Heine-Stieltjes polynomials $y_{k}^{(\zeta)}(u)$
is of order ${\rm Int}[J/2]$ or
${\rm Int}[J/2+1]$,
and always convergent when expanded in terms of $u$
in contrast to the characteristic polynomials
of order $2J+1$ generated from the original eigenvalue problem of (\ref{1}),
where ${\rm Int}[z]$ is the integer part of $z$.
By using (\ref{22}),
the eigen-energies given in (\ref{13}) can also be expressed
as

\begin{equation}\label{24}
E^{(\zeta)}_{ k,n_{1},n_{2},\nu}=2\chi(1+2\delta_{\nu 1})
\left(-b^{(\zeta)}_{1}/b^{(\zeta)}_{0}
+(n_{1}-n_{2})/2\right)\end{equation}
with $J=2k+n_{1}+n_{2}+\nu$, of which the corresponding numerical values
are also provided in the last column of Tables \ref{t1} and \ref{t2}.
It is shown in these Tables
that there is a unique excited state {with $E_{{ k,n_{1},n_{2},\nu}}=0$ and
$\{k={\rm Int}[J/2], n_{1}=n_{2}=\nu\}$ for $J\geq 4$, where
$\nu=0$ when $J$ is even or $\nu=1$ when $J$ is odd.}
Except this unique state, there are many pairs of level energies close to each other, especially
the lowest and the highest a few pairs, when $J$ is small.
With the increasing of $J$, as shown in Table \ref{t2} for example,
more pairs of levels seem almost degenerate.
For example, the difference of excited energies of the ground and the first excited state
$(E^{(1)}_{{ 5,0,1,1}}-E^{(1)}_{{ 6,0,0,0}})/\chi$
is less than $10^{-6}$, though the numerical results up to the third decimal place
shown in Table \ref{t2} are the same. The number of pairs of the almost double-degenerate levels
increases with the increasing of $J$. It is expected that these pairs of level energies become
the same when $J\rightarrow\infty$. The double degeneracy occurs is also due to
time reversal symmetry of the system in the $J\rightarrow\infty$ limit, though it is
not the case when  $J$ is a finite integer.
Since the double degeneracy always occurs for half-integer $J$ cases due to
time reversal symmetry, it can be inferred that the double degeneracy should also occur
for integer $J$ cases after removing the unique level with excitation energy being zero
when $J\rightarrow\infty$ because there will be no difference of integer $J$ cases
from half-integer $J$ cases in the $J\rightarrow\infty$ limit except the unique state in
the integer $J$ case.
Furthermore, with the increasing of $J$,
the level energy distribution of pairs of the almost double-degenerate levels
is symmetric with respect to $E=0$, which is the excitation energy of the unique state,
namely, there are ${\rm Int}[J/2]$ almost doubly degenerate levels with energies $E_{r}>0$
and the same number of pairs of almost doubly degenerate levels with energies  $-E_{r}<0$
for $r=1,~2,\cdots, ~{\rm Int}[J/2]$, which should be helpful in evaluating
the time evolution matrix~\cite{ze} of the system in the large $J$ limit.
Anyway, once the expansion coefficients ${\bf b}$ are obtained,
the results can be used for constructing eigenstates according to
(\ref{exp1}) and (\ref{exp2}), which can then be used
to calculate and analyze physical quantities in the system.

\vskip 1cm
\begin{center}
\textbf{VI. SUMMARY}\vskip.2cm
\end{center}

In this work, by using
the Jordan-Schwinger (differential) boson realization of the SU(2) algebra,
it is shown that the two-axis countertwisting Hamiltonian is exactly solvable
with the help of the algebraic Bethe ansatz
when the quantum number of the total angular momentum of the system $J$ is an integer.
Here, exactly or analytically solvable Hamiltonian means that its entire spectral
problem can be reduced to an algebraic one, which is also related
to its integrability~\cite{ort}. Its solutions can then be obtained
algebraically with eigenvalues expressed in terms
of roots of a set of Bethe ansatz equations.
Though the Hamiltonian for half-integer $J$ case seems also solvable, 
the procedure shown in this work
can not be applied to half-integer $J$ case directly,
which, therefore, has not been addressed in the present study.
It is shown that solutions of the Bethe ansatz equations
can be obtained  as zeros of the Heine-Stieltjes polynomials
determined by the second order Fuchsian type differential equation.
It is verified that the inverse of the zeros are all real and within the two open
intervals $(-1,0)$ and $(0,1)$.
The total number of the four sets of the zeros equals exactly to $2J+1$ for a given
$J$, which proves the completeness of the solutions.
It is also observed that the matrix in determining
the zeros is also tridiagonal and ${\rm Int}[J/2]$ or ${\rm Int}[J/2+1]$ dimensional.
Moreover, there is a non-degenerate unique level with the excitation energy being zero.
It is revealed that there are many pairs of level energies, especially
the first a few lowest and the last a few highest levels, being almost double-degenerate.
The number of the almost double-degenerate levels increases with the increasing of $J$.
Since the double degeneracy always occurs in half-integer $J$ case due to
time reversal symmetry, it can be inferred that the double degeneracy should also occur
in integer $J$ case when $J\rightarrow\infty$ except the unique level.
The level energy distribution of the almost double-degenerate levels
is symmetric with respect to the unique level,
which should be helpful in evaluating
the time evolution matrix~\cite{ze} of the system in the large $J$ limit.
The procedure outlined may be helpful
in calculating physical quantities in the system
in order to produce maximal
squeezed spin states of many-particle systems.

\bigskip

\begin{acknowledgments}
{Support from the  U. S. National Science Foundation
(OCI-0904874, {ACI -1516338}), {U.S. Department
of Energy (DE-SC0005248)}, the Southeastern Universities Research Association,
the China-U. S. Theory Institute for Physics with Exotic Nuclei (CUSTIPEN) (DE-SC0009971),
the National Natural Science Foundation of China (11375080, and 11675071), the
Australian Research Council Discovery Project DP140101492,
and the LSU--LNNU joint research
program (9961) is acknowledged.}

\end{acknowledgments}

\end{document}